\documentclass[fleqn,10pt]{wlscirep}
\usepackage[utf8]{inputenc}
\usepackage[T1]{fontenc}
\usepackage{subfig}
\usepackage{graphicx}
\usepackage{xspace}
\xspaceaddexceptions{]\}}
\makeatletter
\DeclareRobustCommand\onedot{\futurelet\@let@token\@onedot}
\def\@onedot{\ifx\@let@token.\else.\null\fi\xspace}

\def\eg{\emph{e.g}\onedot} 
\def\ie{\emph{i.e}\onedot}

\def\etal{\emph{et al}\onedot}
\makeatother

\title{Oral cancer detection and interpretation: Deep multiple instance learning versus conventional deep single instance learning}

\author[1,*]{Nadezhda Koriakina}
\author[1]{Nata\v{s}a  Sladoje}
\author[2,3]{Vladimir Ba\v{s}i\'{c}}
\author[1]{Joakim Lindblad}
\affil[1]{Centre for Image Analysis, Dept. of Information Technology, Uppsala University, Uppsala 751 05, Sweden}
\affil[2]{Pathology and Cytology Dalarna, County Hospital Falun, Sweden}
\affil[3]{Clinical Research Center Dalarna, Uppsala University, Falun 791 82, Sweden}

\affil[*]{nadezhda.koriakina@it.uu.se}

\begin{abstract}
The current medical standard for setting an oral cancer (OC) diagnosis is histological examination of a tissue sample from the oral cavity. This process is time consuming and more invasive than an alternative approach of acquiring a brush sample followed by cytological analysis. Skilled cytotechnologists are able to detect changes due to malignancy, however, to introduce this approach into clinical routine is associated with challenges such as a lack of experts and labour-intensive work. 
To design a trustworthy OC detection system that would assist cytotechnologists, we are interested in AI-based methods that reliably can detect cancer given only per-patient labels (minimizing annotation bias), and also provide information on which cells are most relevant for the diagnosis (enabling supervision and understanding). 
We, therefore, perform a comparison of a conventional single instance learning (SIL) approach and a modern multiple instance learning (MIL) method suitable for OC detection and interpretation, utilizing three different neural network architectures. 
To facilitate systematic evaluation of the considered approaches, we introduce a synthetic PAP-QMNIST dataset, that serves as a model of OC data, while offering access to per-instance ground truth. 
Our study indicates that on PAP-QMNIST, the SIL performs better, on average, than the MIL approach. Performance at the bag level on real-world cytological data is similar for both methods, yet the single instance approach performs better on average. Visual examination by cytotechnologist indicates that the methods manage to identify cells which deviate from normality, including malignant cells as well as those suspicious for dysplasia.
We share the code as open source at \url{https://github.com/MIDA-group/OralCancerMILvsSIL}
\end{abstract}
\begin{document}

\flushbottom
\maketitle
%
%
\thispagestyle{empty}


\section*{Introduction}

Cancer of the oral cavity and oropharyngeal cancer are on the list of the most common malignancies in the world.
Early detection of cancer is highly desirable as it is a premise of successful treatment. Current gold standard in confirming a diagnosis is histological examination of a tissue biopsy sample, which is time consuming and painful for the patient. An alternative is to develop faster and painless methods 
which rely on taking brush samples from the oral cavity of the patients and a subsequent analysis of the collected and suitably prepared cytological data. A trained cytologist is able to detect abnormalities in samples coming from patients with malignancy by carefully examining the cells in a microscope, however this is a very difficult task; 
a recent study indicates sensitivity and specificity reached by human experts {\it on the patient-level diagnosis} being not more than $80\%$ and $86\%$ in oral cancer cytology screening~\cite{sukegawa2020clinical}. Steps towards improving diagnostic accuracy as well as understanding malignancy better are therefore very valuable.

Automated analysis of cytological data can increase efficiency of the process to a level that opens possibilities for population-wide screening, towards early cancer detection. 
Deep learning (DL) based image classification methods have shown ability to detect differences between malignant versus healthy samples,  without the need for difficult and time consuming labeling of each individual cell~\cite{lu2020deep}. Systems based on such methods for oral cancer (OC) detection could assist cytologists, given that the method is reliable enough. In fact, it is not sufficient for a classifier to provide just an answer 'yes' or 'no' when dealing with the task of cancer detection. In order to trust the classifier's predictions, a human expert needs information about \emph{why} the system reaches a certain decision~\cite{conceiccao2019review}.
However, this becomes a challenging task when only per-patient, and not per-cell, labels are available for training and evaluation, as is typically the case for cytology OC data. In our work, we utilize only relatively few per-cell annotations provided by a cytotechnologist. 
The number of annotated cells is too low for training, thorough evaluation, or reliable interpretation of the DL classification methods, but are still useful for observing similarities and differences between human and machine findings.

Aiming towards efficient and trustworthy cytology-based OC detection, we are interested in DL based methods that are able to learn from only patient-level labels, while still demonstrating both good patient-level performance and also provide information about the individual cells important for a diagnosis, thereby facilitating human interpretation of the decision made by the DL method.

To tackle the challenge, we are following emerging approaches taken in Whole Slide Image (WSI) analysis for (bio)medical applications~\cite{cheplygina2019not}, where labeling of all parts of a WSI is not feasible
and only weak labels are available. More precisely, we are addressing the  OC detection as a weakly supervised learning problem. Due to the large size of WSIs, the  solutions typically involve dividing each WSI into many smaller patches, which are further processed. We utilize this idea too, while still striving to preserve and use holistic, slide level information.

Multiple instance learning (MIL) is a type of weakly supervised learning which can offer methods suitable for OC detection~\cite{quellec2017multiple}.
The core idea of MIL is to treat {\it bags} composed of instances as one object with its well defined label. When MIL is used for binary classification, the aim is to classify negative bags  against positive bags; a bag is classified as positive if it contains at least one positive (key) instance, whereas a negative bag does not comprise any key instances.

WSI analysis with patient-level labels can be seen as an example of MIL, with a set of labeled bags (patient samples), such that a bag is labeled positive if at least one instance (one cell) in it is positive (malignant).
An alternative to MIL is a conventional deep single instance learning (SIL) approach.
In the absence of per-cell annotations, each cell is given a label corresponding to the label of the sample/patient it comes from. 
Since a sample acquired from a patient with a malignancy may include very many normal (non-malignant) cells, to use a patient-level label as an indicator of malignancy on the cell-level 
constitutes a very unreliable label; this approach can therefore be seen as another, differently formulated, weakly supervised learning task.

In this study, we compare two methods, one MIL-based and one  conventional SIL-based approach. The methods are conceptually different, but are both based on deep convolutional neural networks (CNN). 
Our main interest is to: (i) compare the performance of the two end-to-end methods on the per-patient level, as well as on the per-cell level; and (ii) explore the decision-making strategies of both approaches by evaluating the networks' ability to identify the key instances -- the diagnostically most relevant cells.

To be able to reliably evaluate the observed methods, we have created PAP-QMNIST, a synthetic dataset that mimics the main properties of our OC dataset, such as cell image size, colour distribution, arbitrary rotation of cells, amount of blur and noise, number of patients and number of images per each patient. The main   advantage of PAP-QMNIST is that it offers access to reliable ground truth annotation at the instance (cell) level. We envision that the PAP-QMNIST dataset may be useful for several similar studies by other researchers in the field.

The main contributions of our work are:
(i) We introduce a new dataset, PAP-QMNIST, as a model with similar distribution as the cytological data, with well defined ground truth;
(ii) We use PAP-QMNIST to evaluate the performance of two different approaches for weakly supervised OC detection, at patient-level and at cell-level; 
(iii) We compare the two observed approaches on real-world medical image cytological data, focusing on both per-patient and per-cell performance. 

To facilitate reproducibility, we share the complete implementation and evaluation framework as open source.

\section*{Background and related work}
With our main interest in the development of an efficient and reliable automated system for OC detection, we investigate existing weakly supervised methods which are applicable also in cases with a limited amount of labelled data.  This is stated as one of the remaining challenges in computational pathology Lu \etal~\cite{lu2021data}.

Our previous work~\cite{lu2020deep} indicates that it is possible to distinguish cells from healthy patients and cells coming from OC patients, using only per-patient annotations. This method is based on single instance-level training of deep CNN and we use it as a reference method in our study. The same strategy  is evaluated in recent studies, for example in Li \etal~\cite{li2021dual}, where it is referred to as "patch-based without considering MIL".
In our current study, we explore this approach further, to understand whether it can help to find cells that are highly relevant for OC diagnosis. 

There are other methods that can be applied to our problem; they lie on the intersection of two research directions: weak supervision, due to the absence of cell-level annotations, and interpretability, providing cell-level information to reach trustworthy solutions and reliable utilization.
The approaches that can meet both requests and at the same time perform well for WSI analysis are: attention-based deep MIL (ABMIL)~\cite{ilse2018attention}, a recent method by Campanella
\etal~\cite{campanella2019clinical}, dual-stream MIL (DSMIL)~\cite{li2021dual}, clustering-constrained-attention MIL (CLAM)~\cite{lu2021data}, and a more recent transformer-based MIL (TransMIL)~\cite{shao2021transmil}.
The work on ABMIL proposes to incorporate attention mechanisms (and by that interpretability) to MIL, to provide insight into the contribution of each instance. The method by Campanella \etal involves aggregation for MIL based on recurrent neural network (RNN). DSMIL is a fusion of a novel non-local attention-based pooling operator with self-supervised contrastive learning and multiscale pyramidal scheme to extract representations. CLAM is an attention-based method with instance-level clustering.
TransMIL is designed using Transformer~\cite{vaswani2017attention} and considers correlation among instances in a bag.
With our current focus on evaluating feasibility of using weakly supervised learning methods on our OC data from low number of patients, with large number of cells per some patients, and unbalanced number of cells among patients, not all above mentioned methods are equally applicable. Some of the algorithms are partly tailored specifically for histology, and not cytological data analysis: they consider spatial relation among instances (cells), which is not relevant for the rather randomly distributed cells on cytology WSIs. Such are introduction of RNN for aggregation in Campanella \etal~\cite{campanella2019clinical} and multiscale approach in Li \etal~\cite{li2021dual}.

In this study, along with conventional (single instance) DL approach involving training per patch using weak labels, we include a reliable representative of MIL methods -- ABMIL. This method both demonstrates good performance at a bag-level and also offers interpretability at an instance level. ABMIL is an end-to-end method, and so is also the SIL approach we consider; we believe that end-to-end approaches can learn features that represent data better than features obtained  in non-end-to-end manner (\eg engineered or pretrained).
However, ABMIL cannot be applied directly on bags with very large number of instances (such as, \eg,  WSI in cytology) due to GPU memory constrains. Other MIL approaches tackle this problem using, for example, self-supervision~\cite{li2021dual} or features pretrained on ImageNet~\cite{lu2021data}.
In a previous study~\cite{koriakina2021effect} we observed that ABMIL can work well with within-bag sampling, benefiting from end-to-end learning, while overcoming memory constraints imposed by WSI. Therefore, in this study  we evaluate ABMIL with sampling on OC data.

\begin{figure}[t]
 \centering
 \captionsetup[subfigure]{justification=centering}
\subfloat[Images from OC patients]{\includegraphics[width=0.32\textwidth]{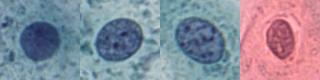}}
\hfil
\subfloat[Images from healthy patients][Images from healthy patients]{\includegraphics[width=0.32\textwidth]{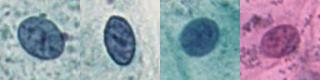}}
\\
\subfloat[Images from a positive bag]{\includegraphics[width=0.32\textwidth]{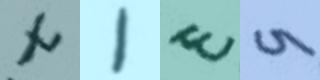}}
\hfil
\subfloat[Images from a negative bag]{\includegraphics[width=0.32\textwidth]{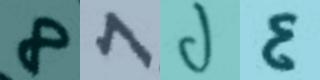}}
\caption{Examples of images from datasets:
\textbf{Top:} OC
\textbf{Bottom:} PAP-QMNIST. Images of the digit ``4'' are key instances in the PAP-QMNIST dataset and appear only in positive bags (example is the leftmost image in (c)), whereas images of other digits appear in both positive and negative bags. }
\label{fig:examples_of_images}
\end{figure}

\section*{Data}
\subsection*{Oral Cancer}
The OC dataset consists of liquid-based (LBC) prepared PAP-stained slides of brush-sampled cells from the oral cavity of 24 patients. Slides were imaged using a NanoZoomer S60 slide scanner, $40\times$, 0.75 NA objective, at 11 z-offsets providing RGB WSIs of size $103936\times107520\times3$, $0.23 \mu m$/pixel.

The method described in Lu \etal~\cite{lu2020deep} was used to detect cell nuclei for which $80\!\times\!80$ pixel regions were cut out, each with one centered nucleus in focus. 
This provides overall $196243$ cells from healthy and $111596$ cells from OC patients.
Example images are shown in Fig.~\ref{fig:examples_of_images}(a),(b). 
We use folded cross-validation where data is separated on the patient level; for each fold
6 healthy patients and 6 OC patients are used for training, 2 healthy patients and 2 OC patients for validation, and 4 healthy patients and 4 OC patients for testing. It always holds that in the test set there are no cells from patients used in training or validation. 
For MIL, we consider all images from one patient as one bag. For SIL, 
images from all patients are considered individually during training and inference, and in the end we aggregate predictions for each patient to compute bag-level performance metrics of SIL in a similar way as we compute them for MIL, taking bag label into account.

The patient-level diagnoses are set based on histological tissue sample analysis (used as ground truth), where ``positive'' labels are associated with patients with a detected malignancy.

\subsection*{PAP-QMNIST}
Reliable cell-level annotations in OC data are scarce. That makes it difficult to judge the outcome of experiments, \eg whether the dataset is too challenging to be processed by a particular method or the parameters of the method are not tuned properly.
To facilitate method development and evaluation, we introduce a new synthetic dataset, PAP-QMNIST \url{https://github.com/MIDA-group/OralCancerMILvsSIL}, 
which is created as a model of OC data while having reliable ground truth also at the instance (cell) level.

Previously, MNIST-bags~\cite{ilse2018attention}, CIFAR10-bags~\cite{seibold2020self}, QMNIST-bags~\cite{koriakina2021effect}, Imagenette-bags~\cite{koriakina2021effect} have been used for MIL experiments. 
These datasets are not well suitable to serve as a model of our OC data; they are not of appropriate size and do not provide enough images per bag (patient) to resemble OC data.

PAP-QMNIST (see Fig.~\ref{fig:examples_of_images}(c),(d)) is created taking these issues into account. We base PAP-QMNIST on the QMNIST dataset~\cite{NEURIPS2019_51c68dc0}, providing a large number of images, and also because the object (digit) is located in the central part of the image, similarly to (the detected) nuclei in our OC data.
We rescale original images to the size of OC images using bilinear interpolation, add colour and augmentations to mimic OC data and replicate the number of patients and number of images per patient in the OC dataset as well. 

Colourisation of resized QMNIST images is performed as follows: from a sample of $5280$ images from the OC dataset, we plot the histogram of each colour channel and approximate them by a normal distribution for the green and blue channels and a skewed normal distribution for the red channel. We then sample colour values from these distribution for each colour channel while avoiding values which, in combination with other two channels, give colourization visually far from colours of OC data.
The same colourisation is applied for positive and negative classes. Knowing, a priori, that the colour is not a relevant feature is useful when evaluating the decision making of the trained networks; if colour appears as an important feature it is an indication that the network has probably over-fitted the training data.

OC data are rotation invariant, therefore, during creation of PAP-QMNIST we rotate digits by a random angle. Additionally, we add horizontal and vertical flips, as well as Gaussian noise, blur, motion blur, random fog and random change in brightness and contrast~\cite{info11020125}.

The original QMNIST dataset is composed of train and test sets, each of 60k images. We randomly extract 20\% of the original QMNIST test set to form a validation set while remaining 80\% are kept as test set. PAP-QMNIST train, validation and test sets are sampled correspondingly from train, validation and test sets of the such split QMNIST.

In our experiments we are evaluating the effect of different ratios of key instances in the data on success of automated detection and interpretation, \ie 
the ratio of OC cells (responsible for the diagnosis) to the total number of cells in a bag
(this being an unknown quantity in the OC data). We therefore dynamically create the datasets with a certain percent of key instances per bag. We choose the digit ``4'' to represent key instances in the PAP-QMNIST data; this provides a suitably difficult problem, where rotated versions of this digit may look very similar to several other digits in the data set. 
The key instances are present \emph{only} within bags representing samples from patients with malignancy.

\section*{Methods}

For both considered methods (ABMIL and SIL), and both datasets (PAP-QMNIST and OC), we 
evaluate several different architectures: LeNet architecture as  in Ilse \etal~\cite{ilse2018attention}, ResNet18~\cite{he2016deep} and SqueezeNet~\cite{iandola2016squeezenet}. The choice of these architectures is made considering the size of images and their preliminary performance on PAP-QMNIST data.

To facilitate reproducibility, we share the complete implementation and evaluation framework as open source \newline \url{https://github.com/MIDA-group/OralCancerMILvsSIL}. 
All experiments are performed using PyTorch~\cite{paszke2019pytorch}. The Albumentations~\cite{info11020125} package is used to augment both PAP-QMNIST and OC data during training, and PAP-QMNIST during creation. Augmentations used during training are horizontal and vertical flips, rotation by multiples of 90 degrees (to avoid interpolation), added Gaussian noise. Images are standardized by mean and standard deviation of the whole dataset, for both OC and PAP-QMNIST. 
Learning rates and weight decay regularisation coefficients were selected for different architectures based on optimization on a subset of PAP-QMNIST. They are (respectively) set as follows: for ABMIL based on LeNet, $5{\,\cdot\,}10^{-5}$ and $10^{-6}$; for ABMIL based on ResNet18, $5{\,\cdot\,}10^{-6}$ and $10^{-5}$, for ABMIL based on SqueezeNet, $5{\,\cdot\,}10^{-5}$ and $10^{-5}$, for SIL based on LeNet, $10^{-4}$ and $10^{-4}$; for SIL based on ResNet18 and SqueezeNet, $10^{-4}$ and $10^{-6}$. 

During training we store the model weights for each epoch. To finally select the model to use for evaluation, we observe the moving average (window of 15 epochs) F1 score for SIL and classification error for ABMIL on the validation set, and for the moving average window with the highest average F1 score/classification error we select the model which provides the best validation F1 score/classification error within that window. F1 score is not used for ABMIL due to the balanced number of bags for two classes.

\subsection*{Attention based deep multiple instance learning with sampling}
We perform experiments using ABMIL with attention mechanism and use the same model as in Ilse \etal~\cite{ilse2018attention} with LeNet based feature extractor. We also evaluate ResNet18 and SqueezeNet as a backbone for feature extraction.
We set the maximum number of training epochs to 1500,
having observed beforehand that the best validation performance is reached well before this number.

We perform experiments with mini-bag sizes with 2500, 1200 and 500 instances, as our previous study indicated increase in Area-Under-Curve (AUC) at the instance level for small mini-bags~\cite{koriakina2021effect}. In addition, lower number of images for sampling reduces GPU utilization.
We use model parallel technique, sending different parts of the model to two different GPUs, each of 32 or 16 GB of VRAM depending on the mini-bag size.

The approximate number of test evaluations for each instance (see Koriakina \etal~\cite{koriakina2021effect}) is calculated based on the bag with highest number of instances, for which we make approximately 10 test evaluations per instance.

\subsection*{Conventional single instance deep learning for classification}
We follow the approach proposed in Lu \etal~\cite{lu2020deep}. We set the overall number of training epochs to 150 for both PAP-QMNIST and OC data, which is defined beforehand observing that the best validation performance is reached  earlier than this number of epochs. We use mini-batch size of 56 images.

\begin{figure*}[t]
  \centering
  \captionsetup[subfigure]{justification=centering}
  \subfloat[][LeNet]{\includegraphics[width=0.3\textwidth,trim={1mm 2mm 1mm 2mm},clip]{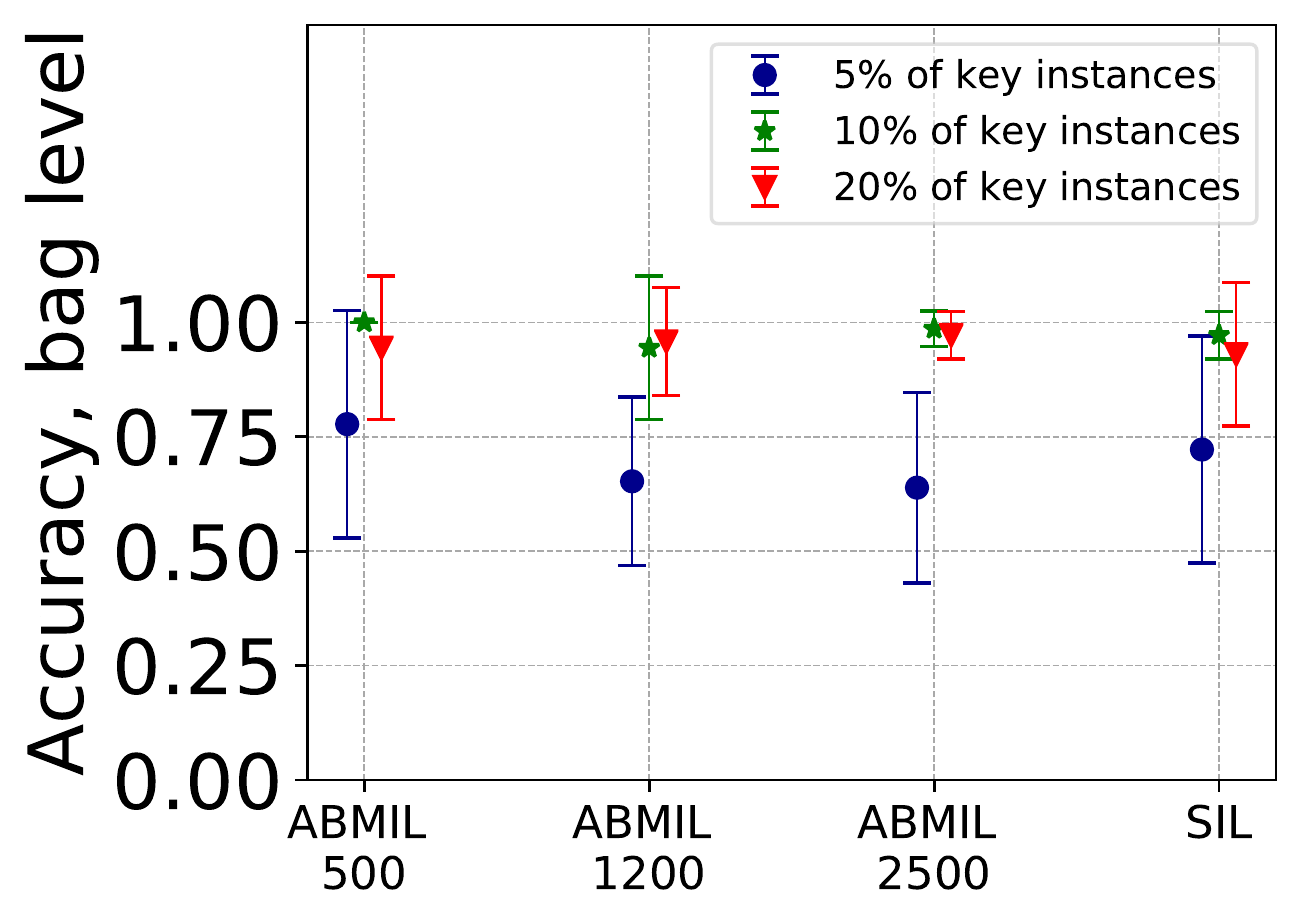}}
 \quad
    \subfloat[][ResNet]{\includegraphics[width=0.3\textwidth,trim={1mm 2mm 1mm 2mm},clip]{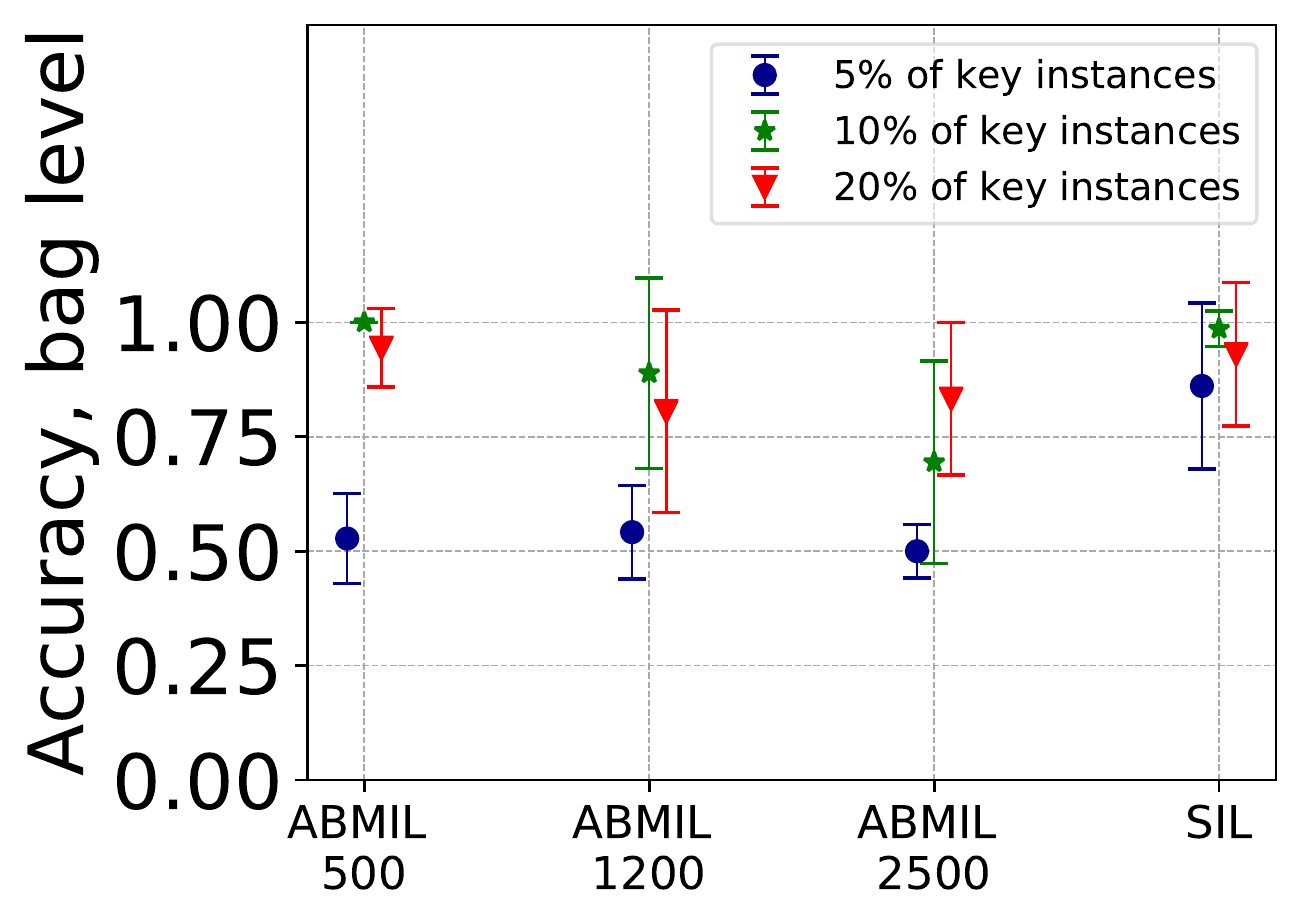}}
 \quad
    \subfloat[][SqueezeNet]{\includegraphics[width=0.3\textwidth,trim={1mm 2mm 1mm 2mm},clip]{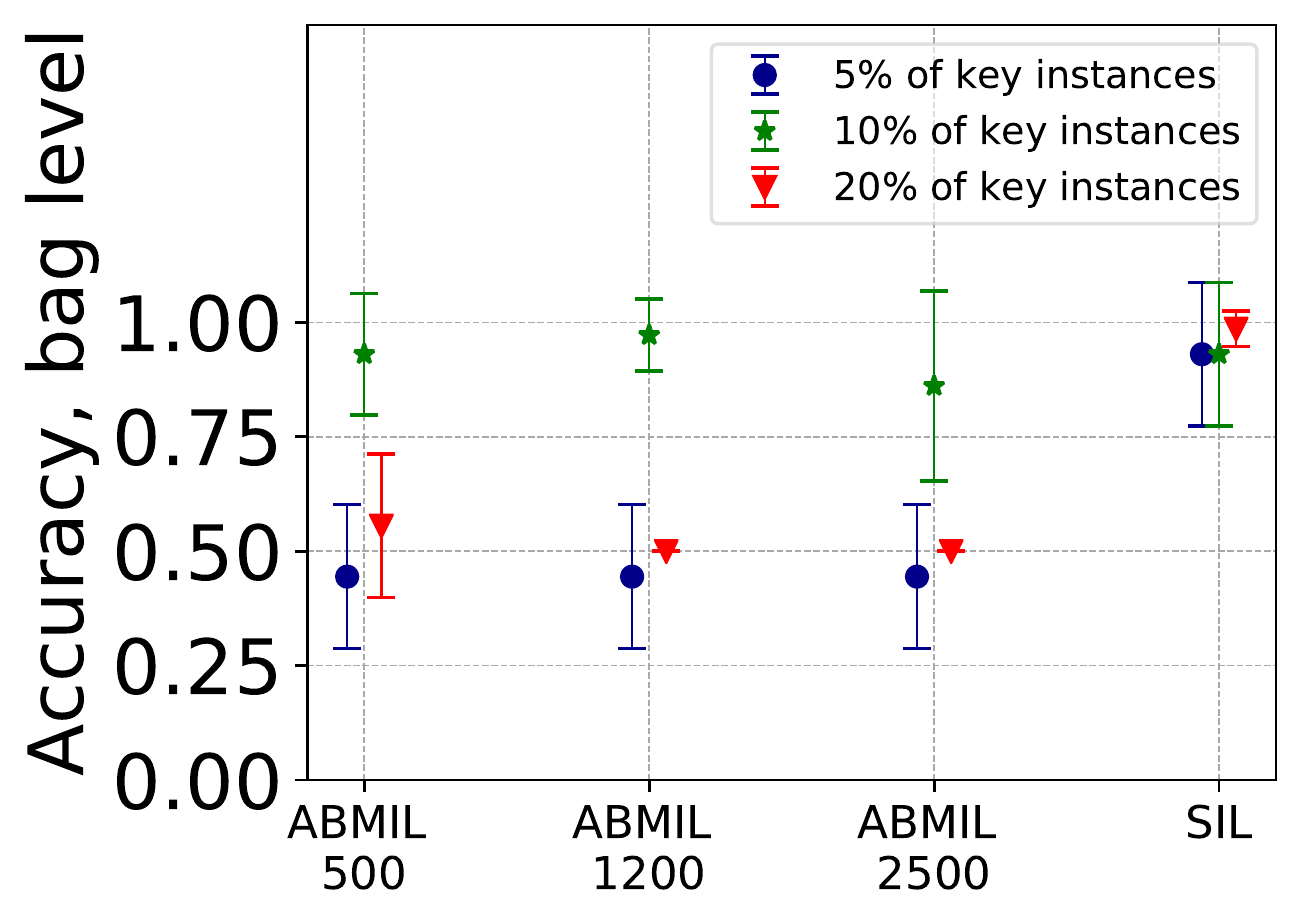}}\quad\newline
  \subfloat[][LeNet]{\includegraphics[width=0.3\textwidth,trim={1mm 2mm 1mm 2mm},clip]{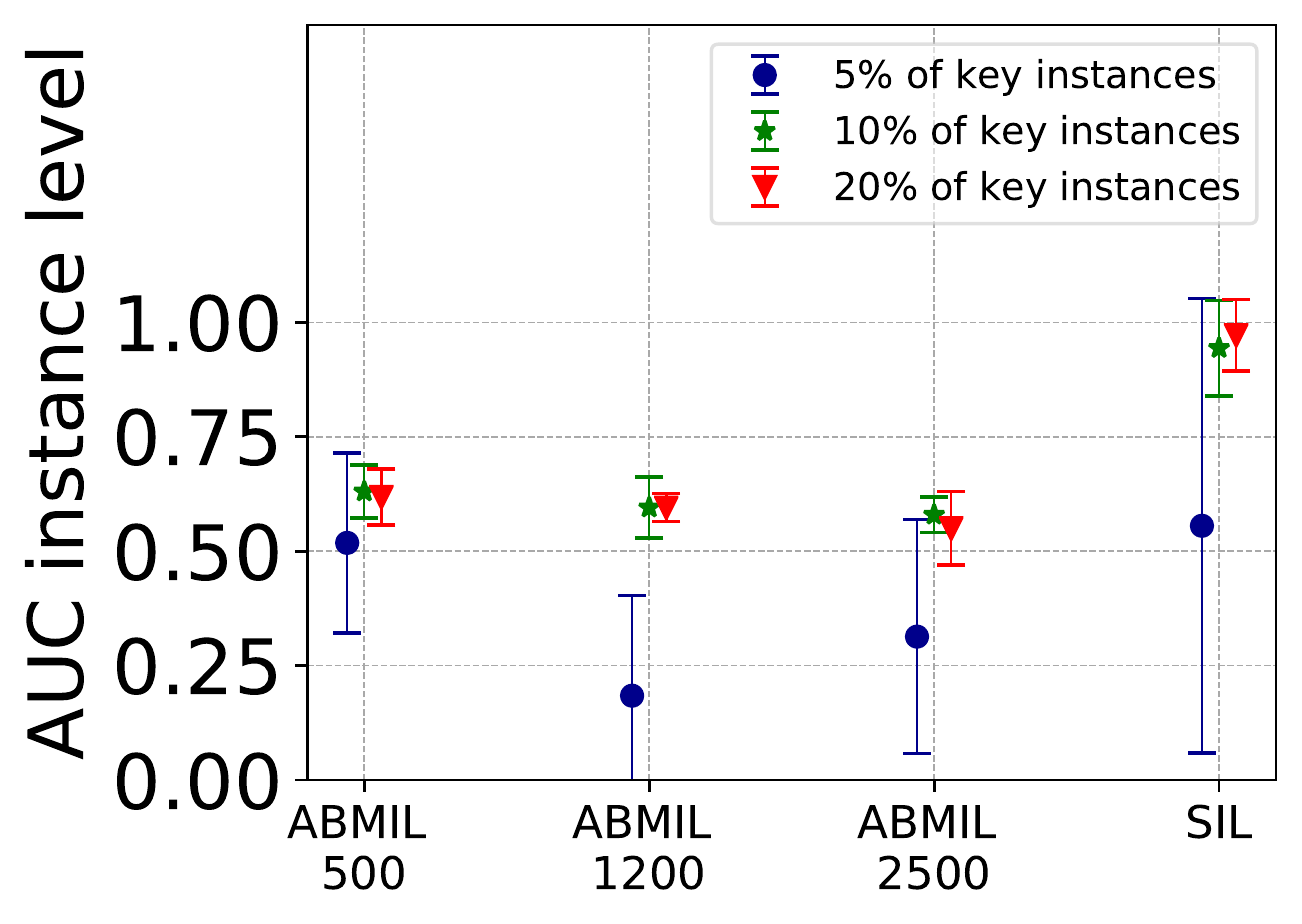}}
  \quad
  \subfloat[][ResNet]{\includegraphics[width=0.3\textwidth,trim={1mm 2mm 1mm 2mm},clip]{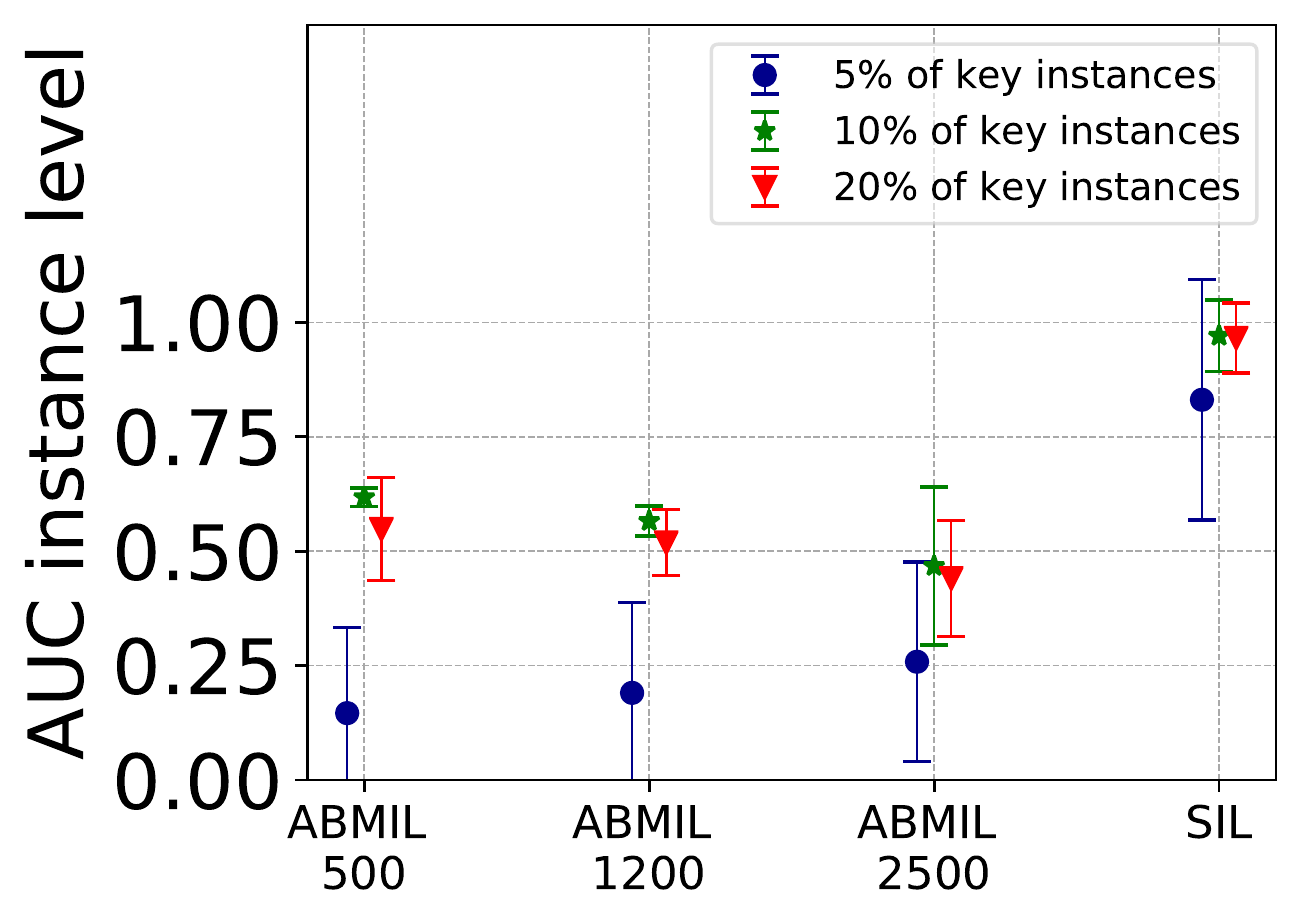}}
  \quad
 \subfloat[][SqueezeNet]{\includegraphics[width=0.3\textwidth,trim={1mm 2mm 1mm 2mm},clip]{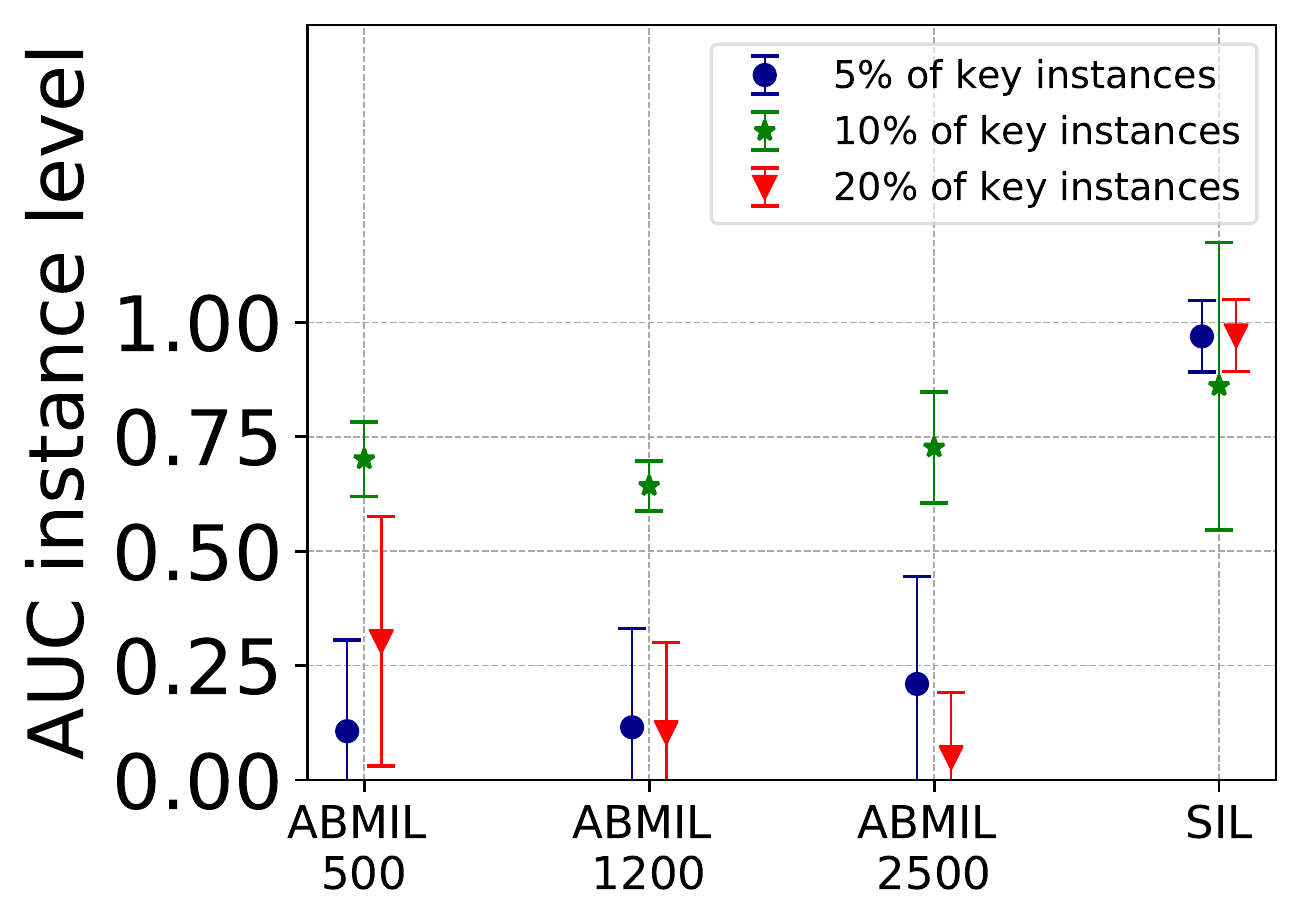}}\quad
\caption{
Average (over 9 folds) test set performance for ABMIL (three mini-bag sizes, 500, 1200 and 2500, are indicated on x axis) and SIL on PAP-QMNIST-bags 
(larger is better):
\textbf{Top:} Accuracy at the bag level.
\textbf{Bottom:} AUC at the instance level. Error bars correspond to standard deviation over 9 folds. 
}
\label{fig:AUC_PAPQMNIST}
\end{figure*}

\section*{Evaluation}
We perform quantitative evaluation of algorithms by computing (1) accuracy at the patient level and the Area under the Receiver Operating Characteristic curve (AUC) at the cell level for PAP-QMNIST and (2) accuracy at the patient level for OC data.

Qualitative evaluation of results on OC data is performed with participation of an expert cytotechnologist. Our aim is to identify if there exists a correlation between the key-instances  identified by the automated system (based on each of the models) and by the human expert. 
The pathologist is provided, for each patient in the test set, the 36 most relevant instances (with highest attention weights/prediction score), as identified by the system in three folds of the OC data, and is asked to 
identify images/cells with characteristics which can be related to malignancy. Cytological annotation was performed in analogy to the Bethesda system~\cite{nayar2017bethesda,sivakumar2021application}, where ASC-US stands for 'atypical squamous cells of unknown significance' and ASC-H for 'atypical squamous cells, cannot exclude a high-grade lesion'.

The images shown to our expert cytotechnologist are cut-outs of a larger size, $170\times170$ pixels instead of the $80\times80$ pixels used by the methods, to make the examination more reliable (since such a setup is more similar to what cytotechnologists typically use).
This difference does not affect the study, 
since we do not compare human performance to the performance of the considered methods, but we are interested in obtaining the opinion of an expert on which cells the methods report as being most related to the malignant class.

One may argue that the evaluated methods could be applied to larger image crops as well.
Even though such a change could possibly improve the overall classification performance, we do not see any reason to believe that this would change the relative performance of the observed methods. Classification of the larger crops are computationally more demanding, especially for the ABMIL method, and, more importantly, larger crops contain more parts of neighbouring cells, which reduces the focus from a particular cell of interest; the cytotechnologist can distinguish clutter from cells and is focusing on the central one, whereas computer methods can take additional clutter as a feature.

\begin{figure}[t]
  \centering
  \captionsetup[subfigure]{justification=centering}
    \subfloat[][10\% key instances per bag, ABMIL]{\includegraphics[width=0.6\textwidth,trim={1mm 2mm 1mm 2mm},clip,angle=-90,origin=c]{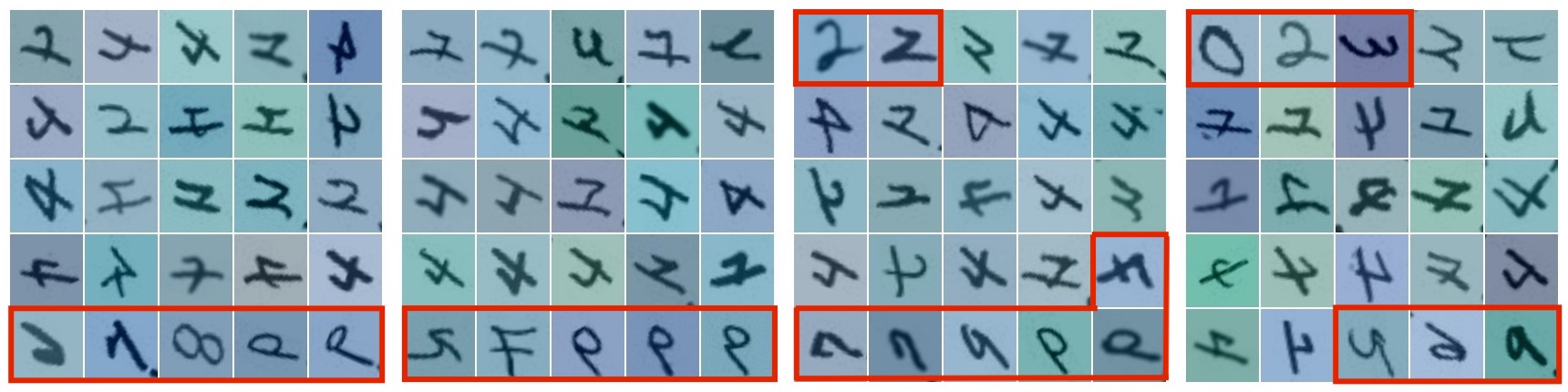}}
 \hfil
    \subfloat[][20\% key instances per bag, ABMIL]{\includegraphics[width=0.6\textwidth,trim={1mm 2mm 1mm 2mm},clip,angle=-90,origin=c]{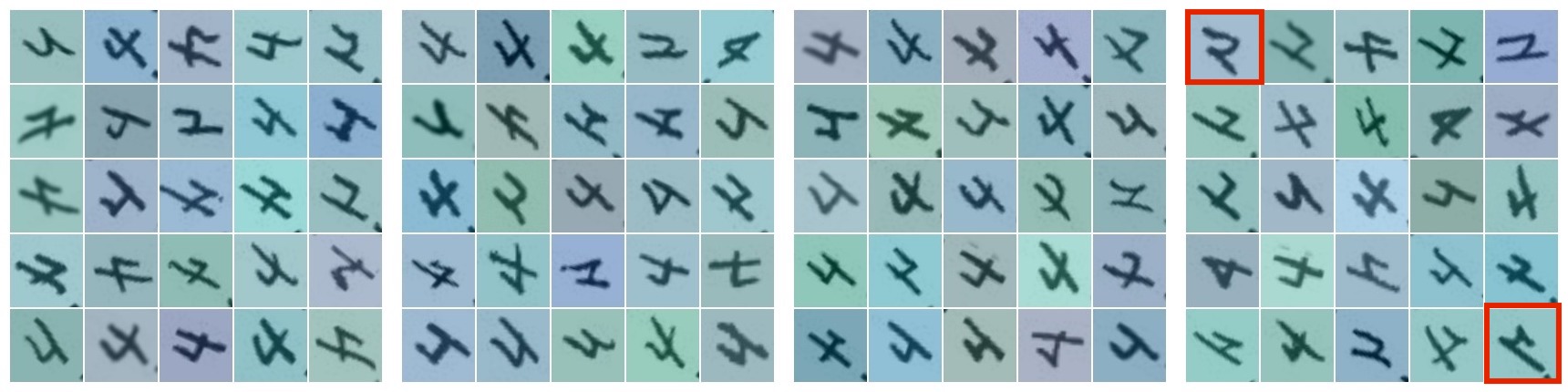}}\hfil
  \subfloat[][5\% key instances per bag, SIL]{\includegraphics[width=0.6\textwidth,trim={1mm 2mm 1mm 2mm},clip,angle=-90,origin=c]{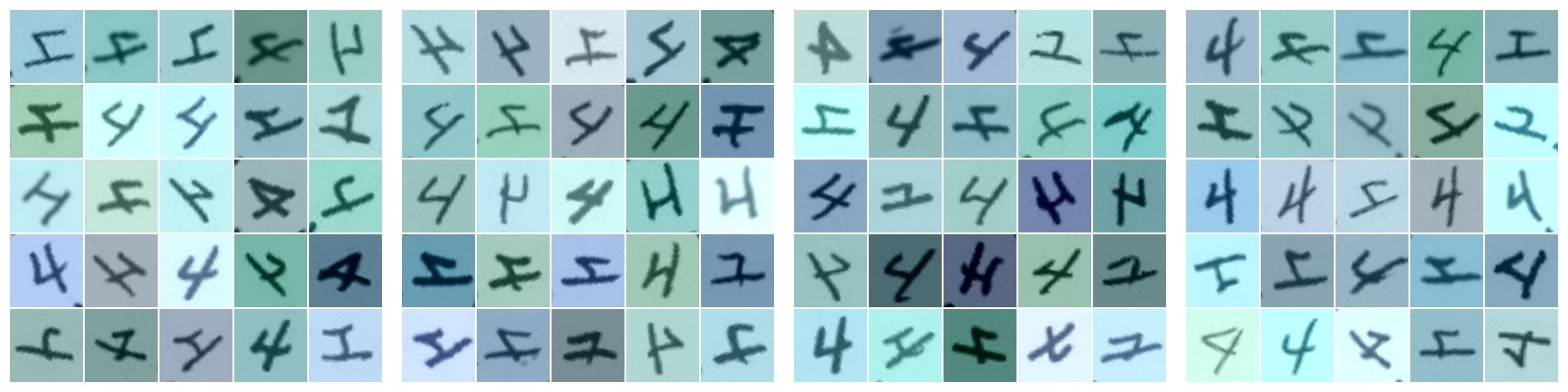}}
  \hfil
  \subfloat[][10\% key instances per bag, SIL]{\includegraphics[width=0.6\textwidth,trim={1mm 2mm 1mm 2mm},clip,angle=-90,origin=c]{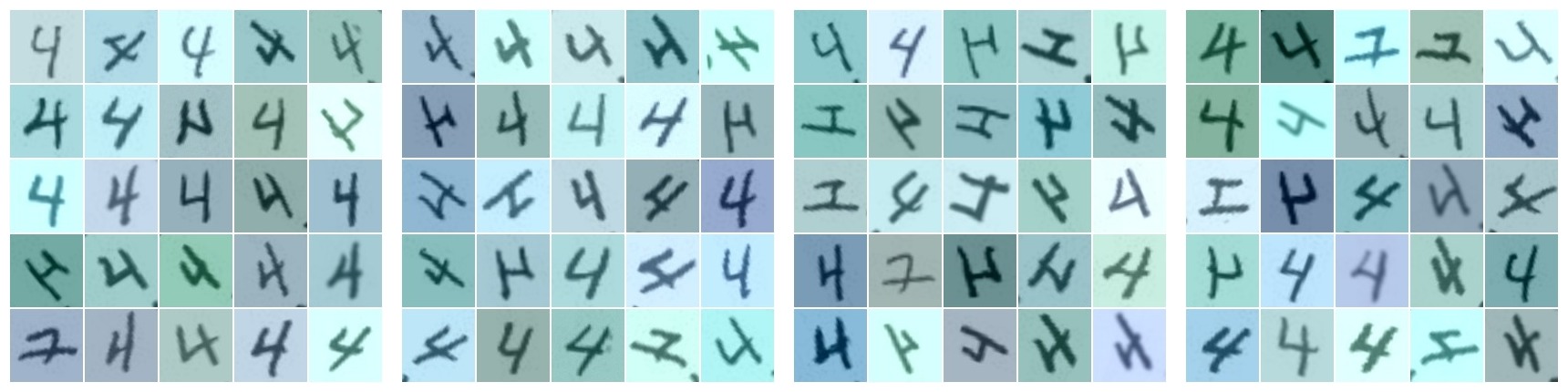}}
  \hfil
 \subfloat[][20\% key instances per bag, SIL]{\includegraphics[width=0.6\textwidth,trim={1mm 2mm 1mm 2mm},clip,angle=-90,origin=c]{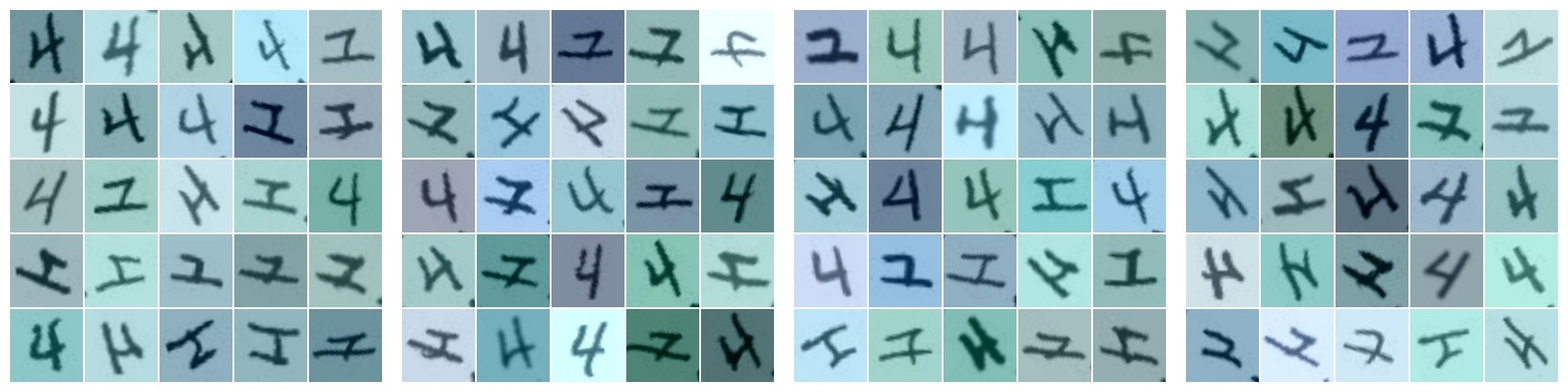}}\quad
\caption{Examples of top 25 instances with the highest attention weight/prediction score for each of the four test bags with positive label, detected by ABMIL and SIL approaches as positive in one of the folds of PAP-QMNIST-bags dataset:
(a)-(b)ABMIL based on ResNet18 and with mini-bag size of 500 instances. ABMIL misclassified the bags with 5\% of key instances, therefore this case is not presented (no key instances are detected).
(c)-(e) SIL based on ResNet18. Red polygons delineate instances 
of digits other than ``4''. 
}
\label{fig:PAPQMNISTkey_ins}
\end{figure}

\subsection*{Attention based deep multiple instance learning with sampling}
We calculate accuracy at the bag (patient) level in the same manner as in Koriakina \etal~\cite{koriakina2021effect}, by computing a majority voted mini-bag label for each test bag. 
AUC at the instance level is calculated only for the test set of PAP-QMNIST, following the same strategy as in Koriakina \etal~\cite{koriakina2021effect}, to find an aggregated attention weight for each instance: we store attention weights for an instance from all mini-bags where this instance is present and then compute the average attention weight using only weights coming from mini-bags with the majority bag label for the instance over all mini-bags this instance is present. The weights are normalised for each mini-bag to the range [0,1], and thresholds in that range are used to compute instance-level AUC. A key instance is detected correctly when the average attention weight for an instance is above the threshold, the majority predicted bag label for an instance is positive and the true instance label is positive, meaning it is a true key instance.

\begin{figure}[t]
 \centering
 \captionsetup[subfigure]{justification=centering}
  \subfloat[LeNet]{\includegraphics[width=1\textwidth]{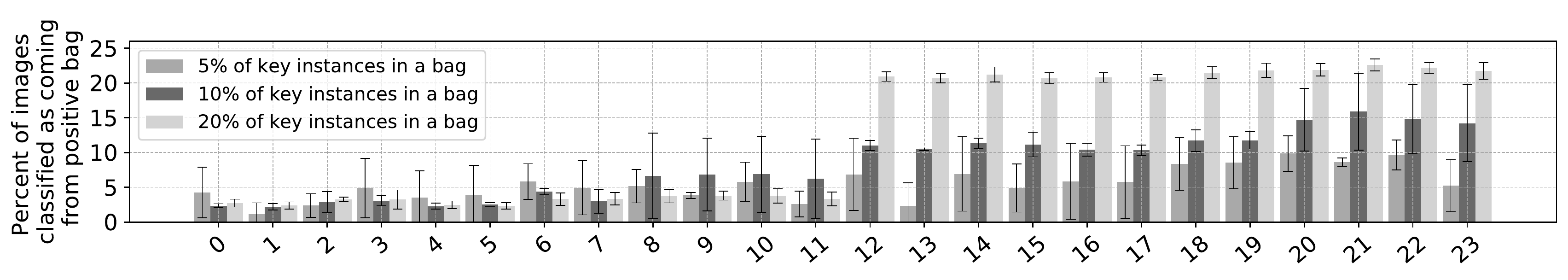}}
  \hfil 
 \subfloat[ResNet18]{\includegraphics[width=1\textwidth]{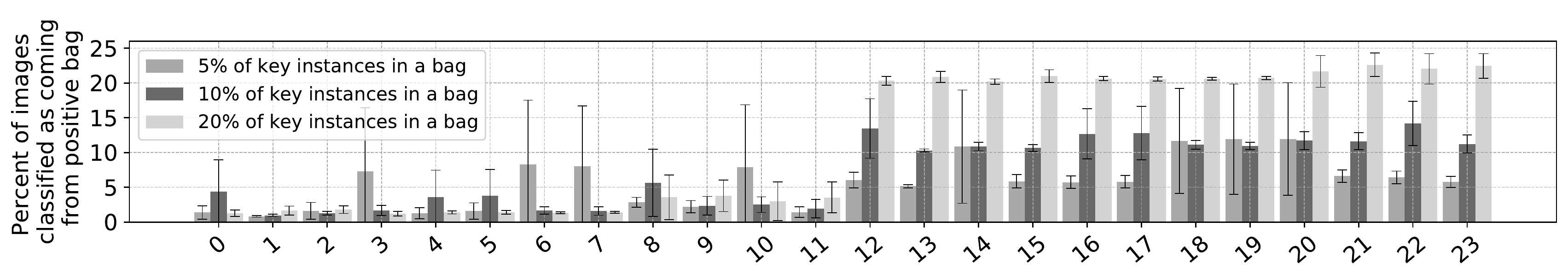}}
 \hfil 
 \subfloat[SqueezeNet]{\includegraphics[width=1\textwidth]{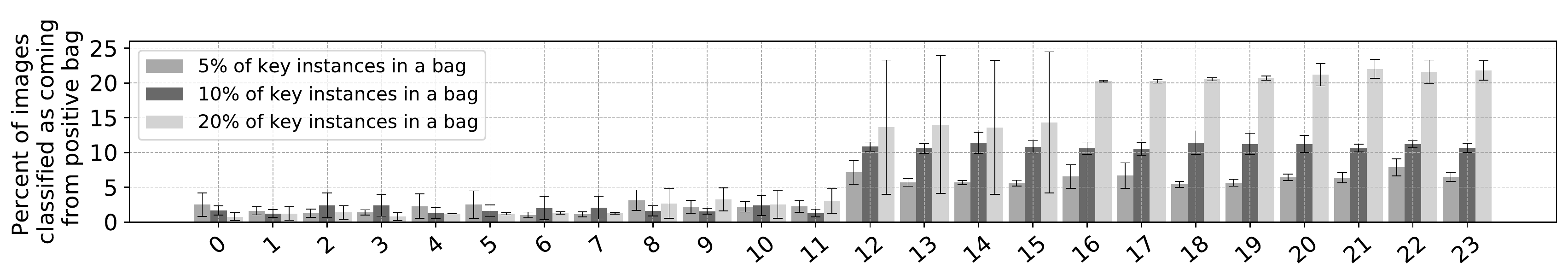}}
\caption{Percentage of cells identified as positive of PAP-QMNIST dataset, per each test bag, by SIL approach. Error bars correspond to the 3 different folds in which each bag appears. Bags 0-11 are negative, bags 12-23 are positive.}
\label{fig:perc_pos_per_bagPAPvanilla}
\end{figure}

\subsection*{Conventional deep single instance learning for classification}
We compute accuracy at the bag level for SIL approach by aggregating the instance predictions for each bag
where the threshold separating positive and negative bags is the average of thresholds from the 9 validation folds computed as the middle value between the percentage of images classified as positive in positive and negative bags.
We compute AUC at the instance level for test PAP-QMNIST set by using prediction scores from the softmax layer.

Due to the different nature of MIL and SIL, bags classified as negative by SIL may contain 
instances which are classified as positive,
whereas bags classified as negative by ABMIL 
may, by definition, not contain any positive (\ie. key) instances~\cite{ilse2018attention}. Therefore, for the sake of methods comparison, we compute key instance detection
performance of SIL in a similar way as for ABMIL, taking into account whether an instance comes from a bag predicted as positive or not.

\section*{Experiments and Results}

We use 9-fold cross validation for all methods and datasets. Each bag appears in the test set of three folds with different combinations of bags for training, validation and test.

\subsection*{PAP-QMNIST}
When creating PAP-QMNIST experimental data, we can define the number of key instances in a bag (patient sample). Ideally, this number should mimic the number of key instances for a patient in the OC dataset. However, it is virtually impossible to estimate the percentage of key instances in the OC dataset. Therefore, we make experiments with different versions of PAP-QMNIST and we set the percentage of key instances for each bag to 5, 10 or 20. We assume an ideal case with the same percentage of key instances for all patients. Bags in folds of PAP-QMNIST are sampled once and then permuted for different folds.

Accuracy at the bag level and AUC at the instance level for all considered methods are shown in Fig.~\ref{fig:AUC_PAPQMNIST}. On average, for PAP-QMNIST, metrics have higher values with 10\% and 20\% of key instances per positive bag, than 
with 5\% of key instances per bag, for all the network architectures used. We also observe that the training is more sensitive to hyper-parameter tuning at 5\% key instances.
Accuracy and AUC values of SIL approach are higher or comparable to those of ABMIL. SIL is able to detect the majority of the key instances for datasets with 10 and 20\% of key instances.

Examples of instances with the highest attention weight/prediction score, for PAP-QMNIST dataset with different percentage of key instances, from bags classified as positive by ABMIL and SIL are presented in Fig.~\ref{fig:PAPQMNISTkey_ins}. Test bags from PAP-QMNIST dataset with positive label and with 5\% of key instances were classified as negative by ABMIL (based on ResNet18 and with mini-bag size of 500 instances), and are not shown in Fig.~\ref{fig:PAPQMNISTkey_ins}.
One can observe that more diverse in colour and rotation key instances are detected by SIL as compared to ABMIL, as well as that a higher number of true key instances within the instances with the highest attention weight/prediction score are found by SIL than by ABMIL.

Fig.~\ref{fig:perc_pos_per_bagPAPvanilla} illustrates the percentage of instances detected by SIL as coming from a positive bag for different architectures and percentage of key instances per bag. The percent of instances classified as coming from a positive bag roughly corresponds to the true percent of key instances per bag, supported by the high AUC at the instance level in Fig.~\ref{fig:AUC_PAPQMNIST}(d-f).

\subsection*{OC}
\begin{figure}[t]
\centerline{\includegraphics[width=0.37\textwidth]{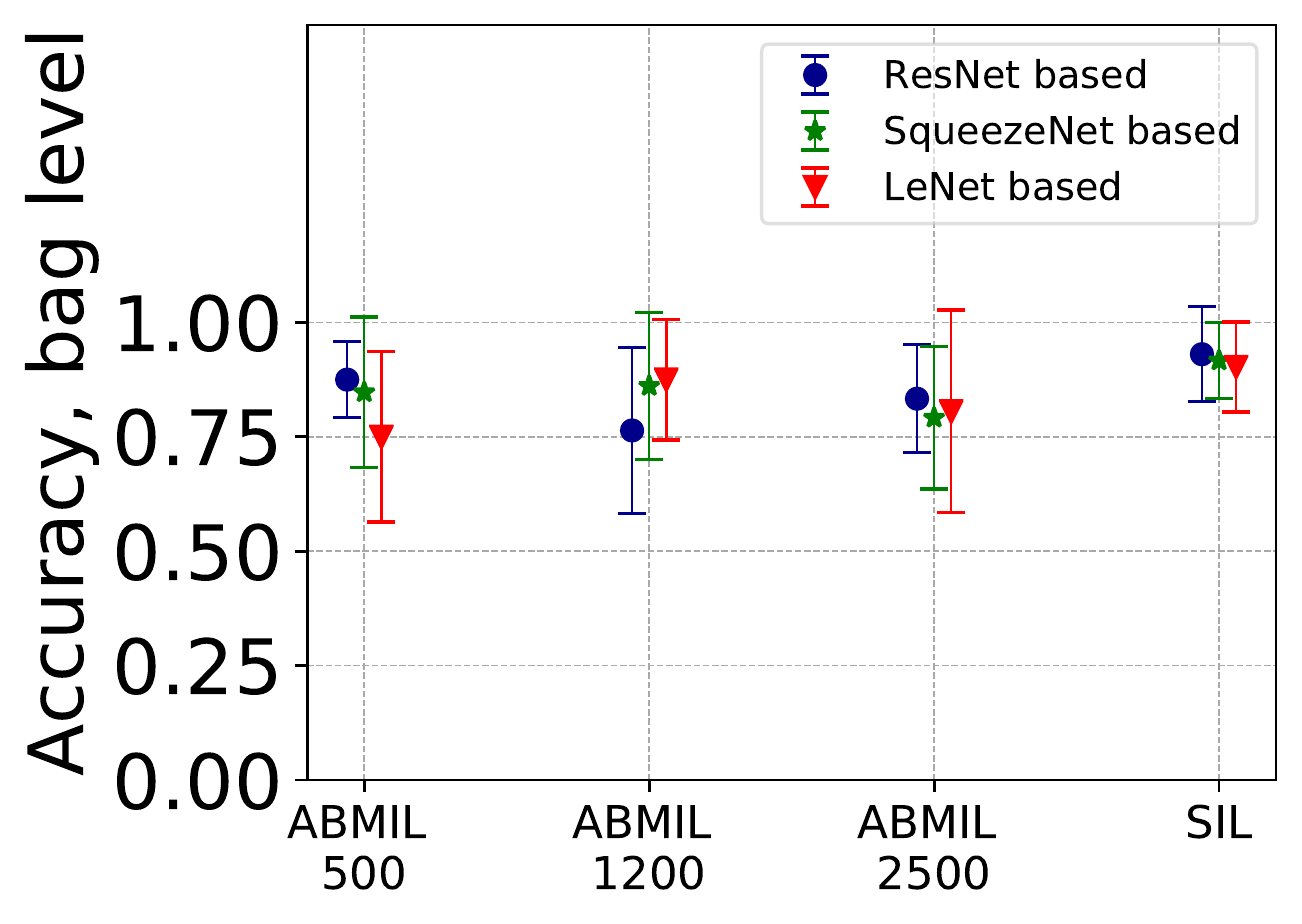}}
\caption{Accuracy at the bag level (averaged over 9 folds) of the ABMIL (three mini-bag sizes, 500, 1200 and 2500, are indicated on x axis) and SIL methods on OC test data. Error bars correspond to standard deviation of accuracy. }
\label{fig:AUCbag_OC}
\end{figure}

\begin{figure*}[t]
\centering
 \captionsetup[subfigure]{justification=centering}
 \subfloat[LeNet]{\includegraphics[width=0.55\textwidth,trim=2mm 0 0 0,clip]{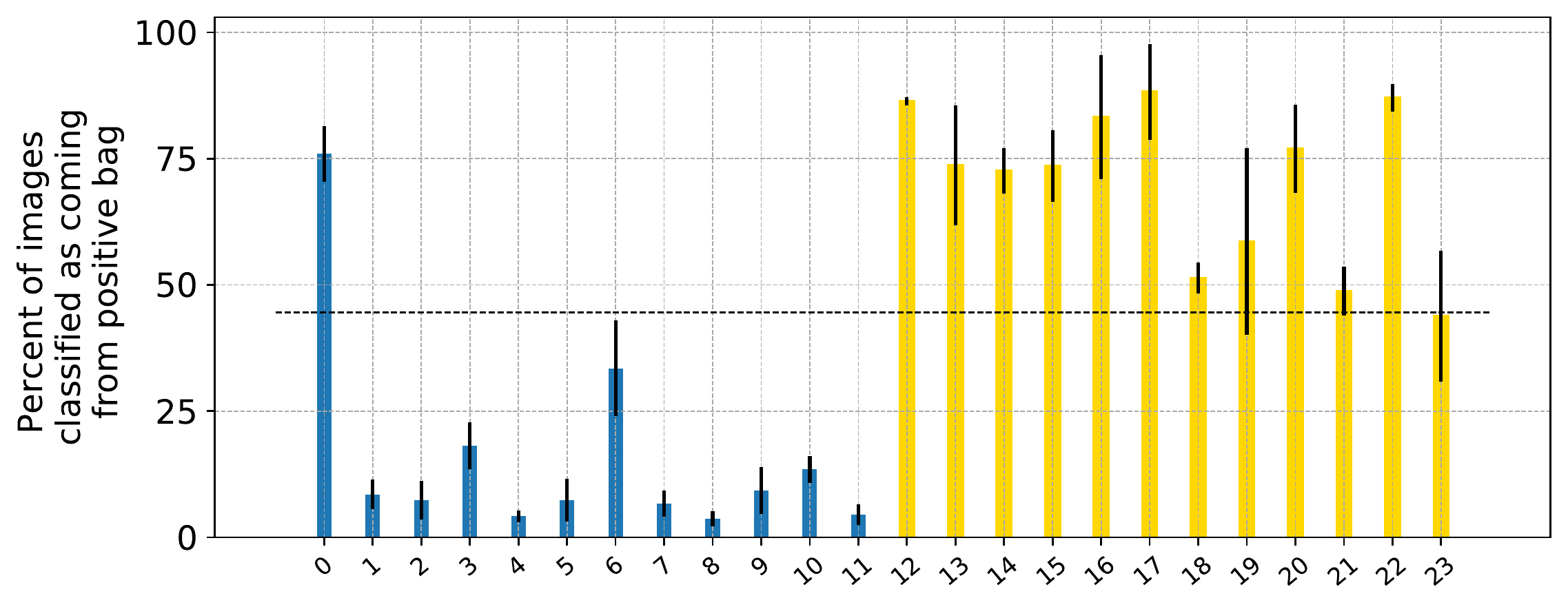}}
 \hfil
 \subfloat[ResNet18]{\includegraphics[width=0.55\textwidth,trim=2mm 0 0 0,clip]{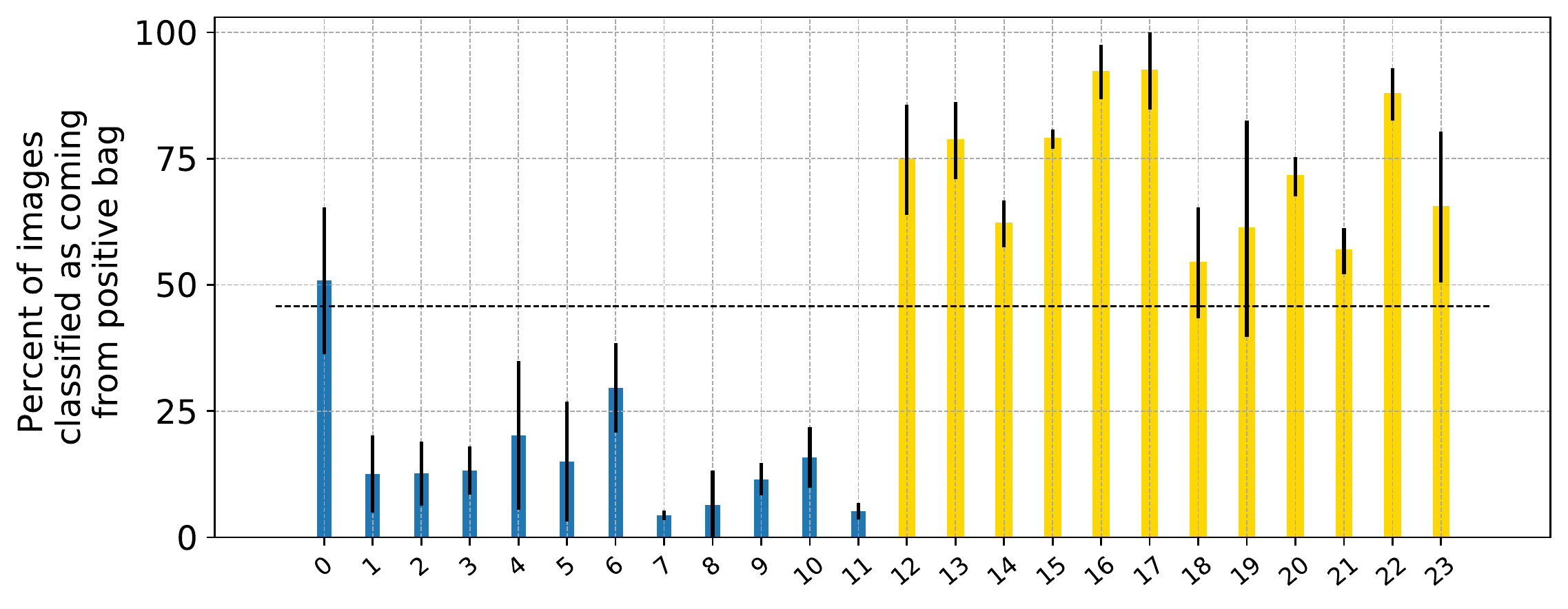}} 
 \hfil
 \subfloat[SqueezeNet]{\includegraphics[width=0.55\textwidth,trim=2mm 0 0 0,clip]{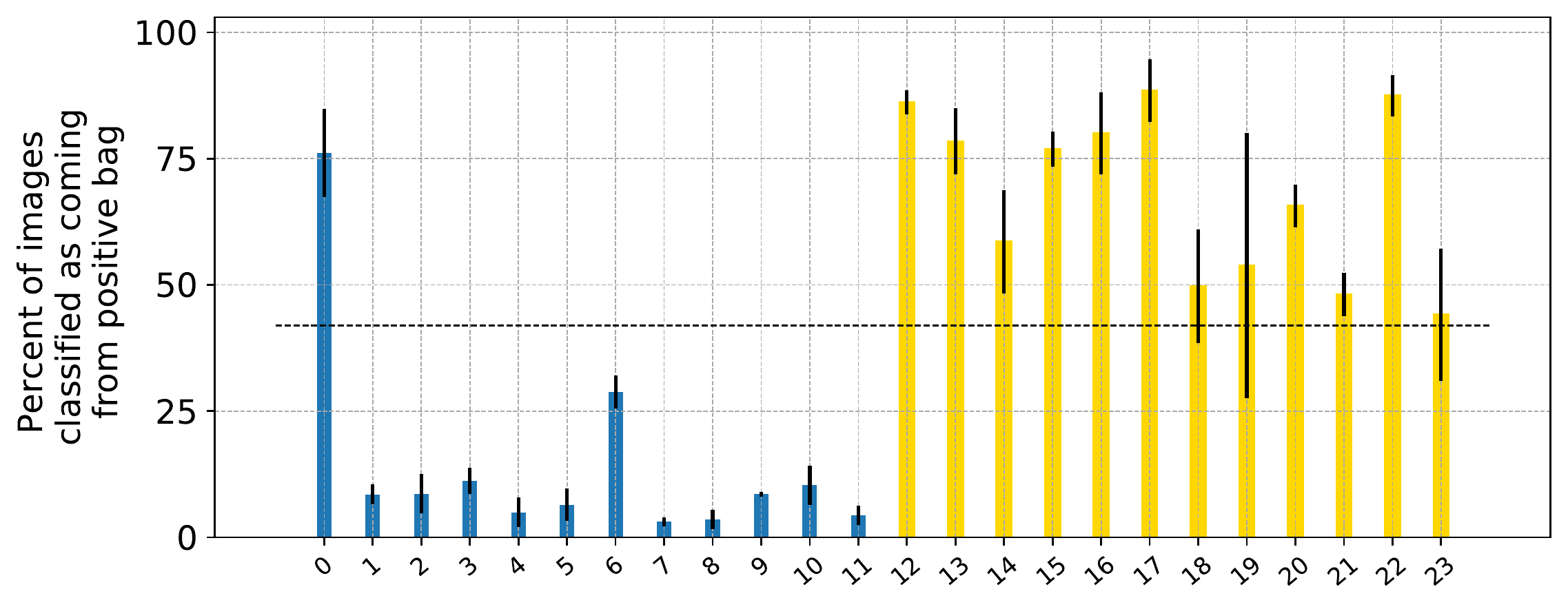}}
\caption{Percentage of cells identified as coming from positive bags of OC dataset, per each test bag, by SIL approach. Error bars correspond to the 3 different folds in which each bag appears. The black dashed line is the threshold computed based on validation set over 9 folds for each architecture type. Bags 0-11 (blue) are negative, the rest (yellow) are positive.}
\label{fig:perc_pos_per_bagOCvanilla}
\end{figure*}

Accuracy at the bag level for OC data is shown in Fig.~\ref{fig:AUCbag_OC}, where accuracy of methods based on considered architectures, on average, are above 0.75. Accuracy of SIL is slightly higher than accuracy of ABMIL. Average accuracy exceeds 0.9 by SIL based on all three architectures.
The percentage of cells identified by SIL as coming from patients with OC for each of the 24 slides, is illustrated in Fig.~\ref{fig:perc_pos_per_bagOCvanilla}.

\begin{figure*}[t]
  \centering
  \captionsetup[subfigure]{justification=centering}
    \subfloat[][Predicted as positive by ABMIL, annotated as normal superficial cells by cytotechnologist]{\includegraphics[width=0.23\textwidth,clip]{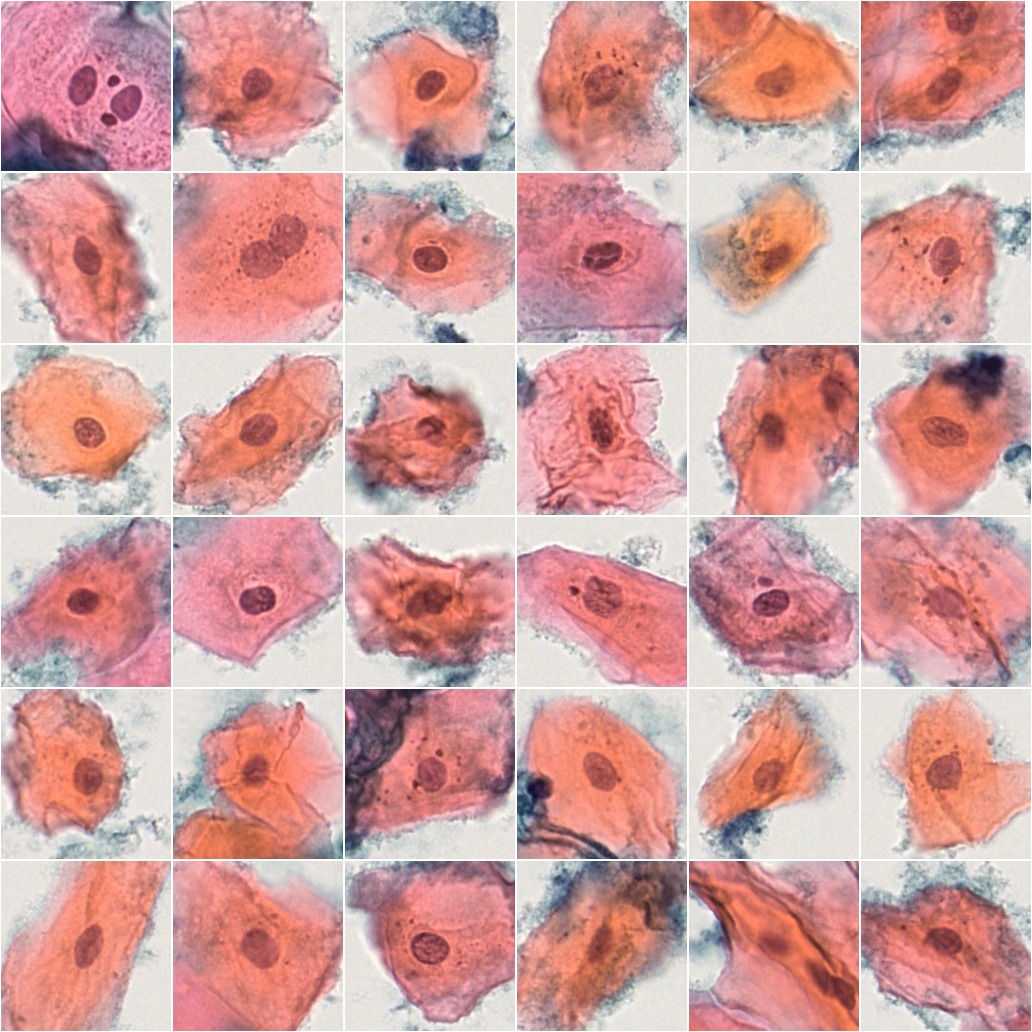}}\quad
  \subfloat[][Predicted as positive by ABMIL, several cells annotated as ASC-US by cytotechnologist]{\includegraphics[width=0.23\textwidth,clip]{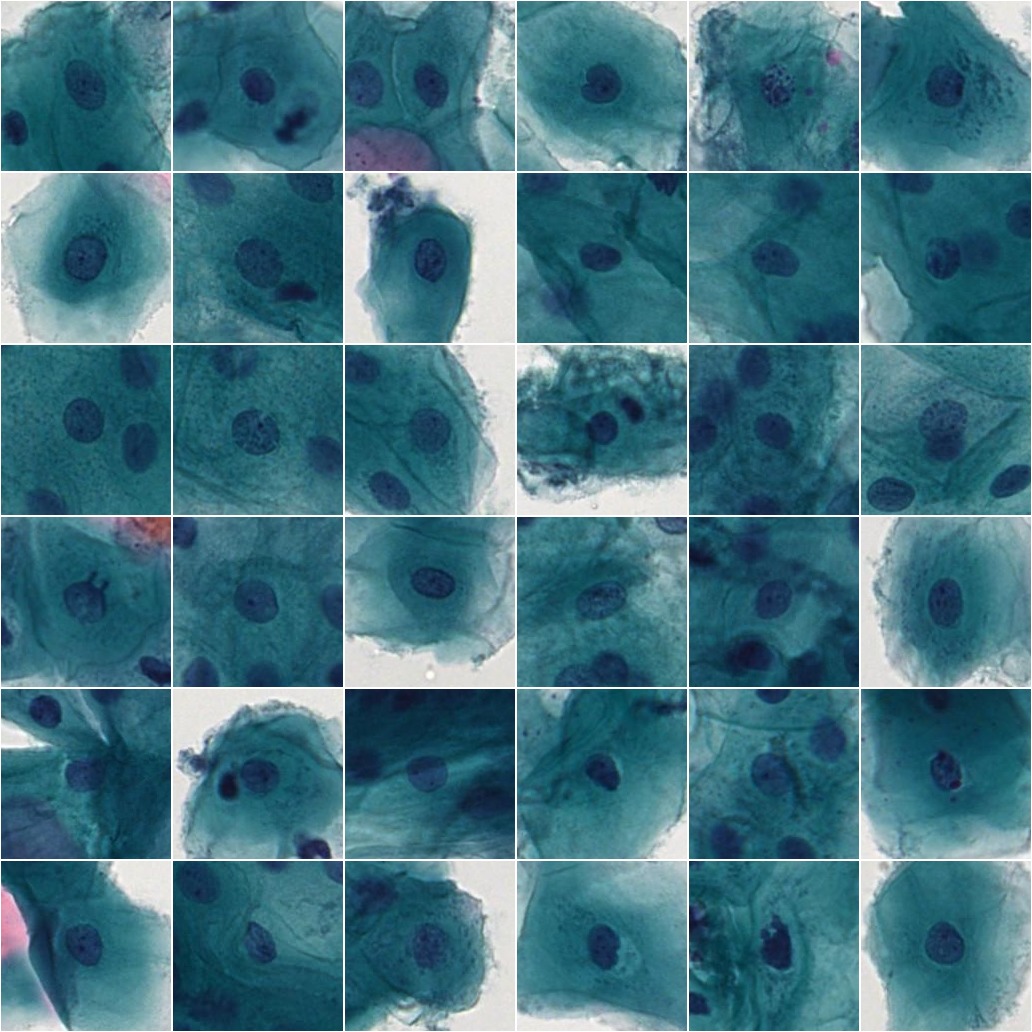}}
 \quad
    \subfloat[][Predicted as positive by SIL, several cells annotated as suspicious for high grade dysplasia by cytotechnologist]{\includegraphics[width=0.23\textwidth,clip]{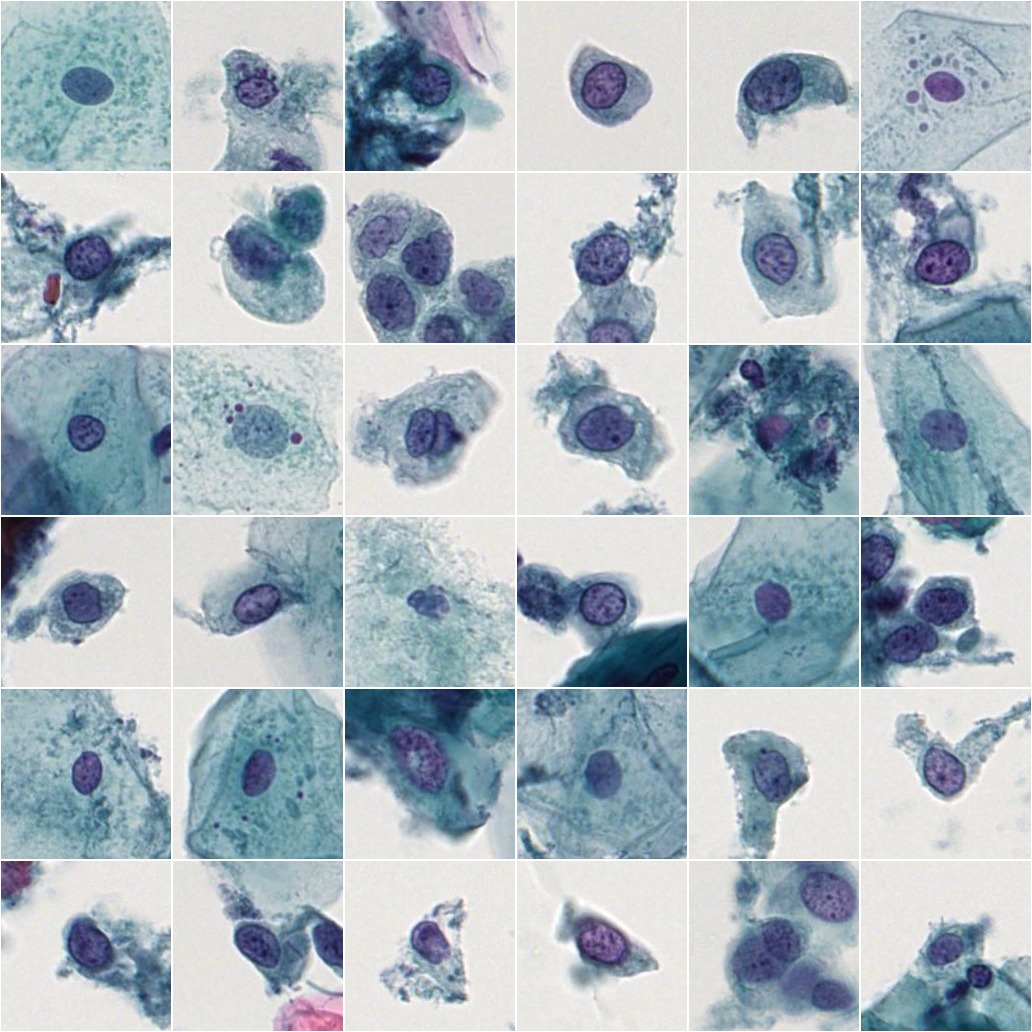}}
  \quad
    \subfloat[][Predicted as negative by SIL, annotated as normal intermediary cells by cytotechnologist]{\includegraphics[width=0.23\textwidth,clip]{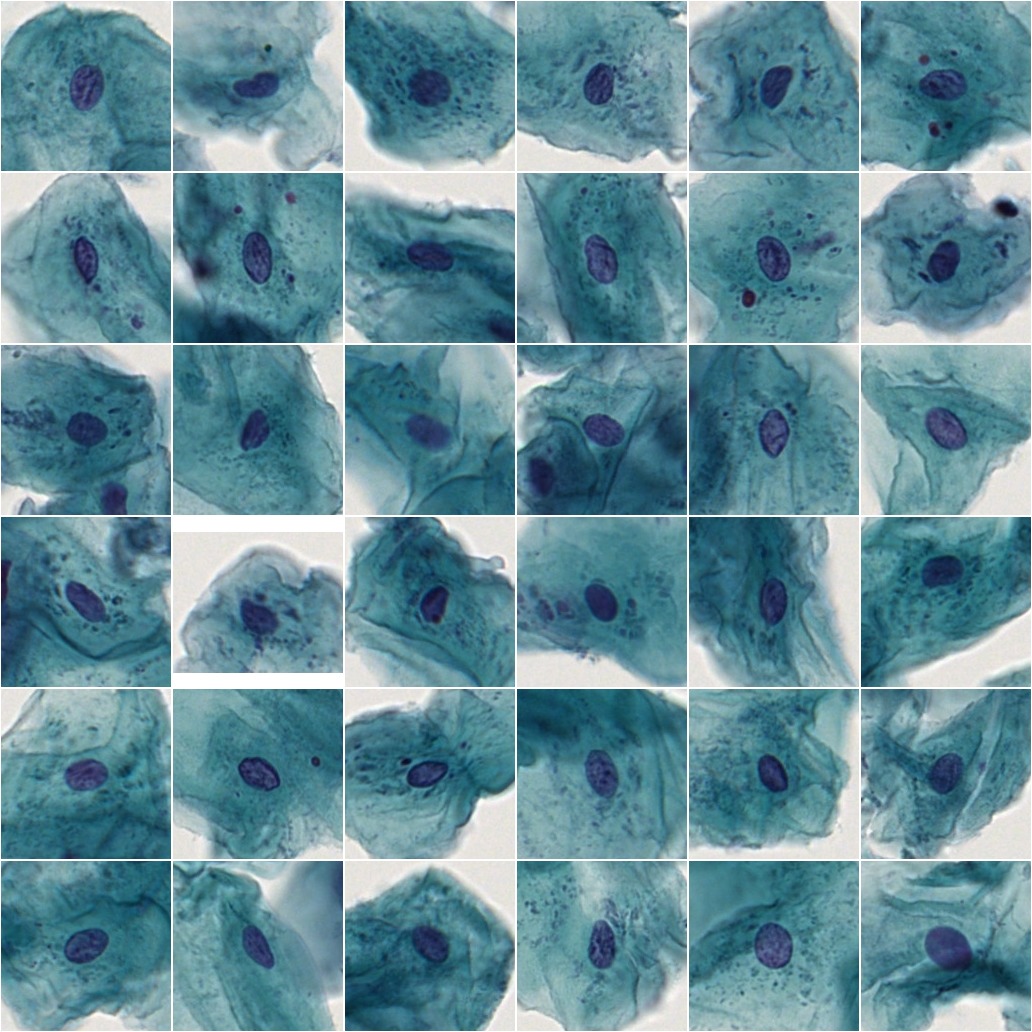}}
 \quad
\caption{Examples of instances with the highest attention weight/prediction score in OC test set found by ABMIL and SIL approaches and corresponding annotations from cytotechnologist.}
\label{fig:key_ins_examples_OC}
\end{figure*}

Figure~\ref{fig:key_ins_examples_OC} shows four mosaics of 36 cells each, in a form presented to the cytotechnologist. We prepared 117 such mosaics. Each selection of 36 cells contains the instances (out of, on average, 9300 cell per patient) with the highest attention weight/prediction score from the bags predicted as positive by the methods.
93 of these mosaics contained no suspicious cells as annotated by the cytotechnologist, but only normal superficial cells (see Fig.~\ref{fig:key_ins_examples_OC}(a)), a mixture of normal superficial cells and normal intermediary cells, normal intermediary cells, or a mixture of normal superficial cells and cell debris or such blood cells as lymphocytes and neutrophils;
21 of the sets were annotated as benign but with reactive changes,  (Fig.~\ref{fig:key_ins_examples_OC}(b)) or sets that contained a minority of not normal cells, such as metaplastic, ASC-US, or ASC-H, with the rest of the cells being normal. 
Three of the sets were annotated as suspicious for high grade dysplasia (see Fig.~\ref{fig:key_ins_examples_OC}(c)), all coming from one patient and detected by SIL with three architectures. For reference, a set of cells predicted by SIL approach as negative and annotated as normal cells are shown in Fig.~\ref{fig:key_ins_examples_OC}(d).
Out of the mosaics with the top-36 key instances detected by ABMIL from bags predicted as positive, 18\% contained cells annotated as reactive or not normal, while of the mosaics with key instances detected using SIL, 28\% contained cells annotated as reactive or not normal.

\section*{Discussion and Conclusion}
In this study we investigate two different methods with the aim to reach interpretable classification of weakly supervised OC data: ABMIL with within-bags sampling, an approach that belongs to the MIL family of methods, and conventional SIL classification. For each method we evaluate three different base architectures, LeNet, ResNet18, and SqueezeNet. 
We evaluate performance on PAP-QMNIST -- an artificial dataset mimicking our OC cytological data, as well as real OC data from 24 patients. 

For ABMIL we observe increased accuracy at the bag level and AUC at the instance level when the percentage of key instances in positive bags increases. We see that high accuracy at the bag level corresponds to high AUC at the instance level for this method. To summarise, we are able to detect key instances from PAP-QMNIST when the percent of key instances in positive bags is not less than 10. ABMIL with mini-bags of size 500 instances, on average, outperformed ABMIL with larger mini-bags. 
The theory of MIL defines that one key instance in a bag makes the bag positive, however, we observe when training on PAP-QMNIST (with number of bags and instances in bags corresponding to what is present in OC data) that one key instance is not enough for ABMIL to classify bags reliably, but rather some percentage of key instances per positive bag is required.

The SIL approach, on average, outperformed ABMIL at the bag and instance levels on PAP-QMNIST data, with ResNet18 and SqueezeNet reaching high values of accuracy and AUC,  also for datasets with (only) 5\% of key instances.
Studying the key instances detected by ABMIL in the PAP-QMNIST dataset (Fig.~\ref{fig:PAPQMNISTkey_ins}) indicate that ABMIL can be prone to paying attention to colour and orientation, features which for this dataset are know (by design) to be not relevant.

Accuracy at the bag level, averaged over 9 folds,  of the different methods on real-world cytological data is lower than accuracy at the bag level of the synthetic PAP-QMNIST,
however, on average surpassing 0.75 and with the SIL approach reaching past 0.9 for all three considered architectures. 

Evaluation of methods on OC data at the instance level is performed involving an expert. The majority of sets of instances with the highest attention weight/prediction score from the bags predicted as positive by the methods are annotated by the cytotechnologist as normal superficial cells or cells/debris not related to malignancy. Such superficial cells have a distinct orange/pink colour and are not specific for malignancy but can be present both within healthy patients and within patients with OC. A possible explanation why such cells are predicted as key instances is the fact that they appear more frequently within patients with malignancy for our current dataset than within healthy patients. 
Also, among sets of predicted key instances there are cells which do not indicate clear malignancy but rather show some deviation from normality. Such cells are detected by both ABMIL and SIL, and are relevant for early detection of OC, \eg suggesting further investigation. The SIL approach was able to detect cells from a patient with high grade dysplasia, which was not observed among the top 36 key instances using ABMIL for the same annotated data. 
Note that we do not consider the visual inspection of only 36 cells out of ca 10,000 cells on a whole slide to correspond to any envisioned realistic use case, but is a means for us to compare the methods.

An overall conclusion of the performed comparative study is that for this task we do not observe any reasons to use ABMIL instead of the less complex and less memory demanding SIL approach. 
This stands in contrast to a common assumption that MIL is more suitable for learning from weakly annotated data that SIL.
Possibly our result is due to the nature of the here addressed task, with rather few and very large bags of varying sizes. 
By the performed experiments we demonstrate that it is possible, not only to separate healthy patients from patients with malignancy, but also to detect cells related to malignancy by utilizing only patient level annotations on a fairly small number of patients. Furthermore, obtained results were consistent on the two datasets. The PAP-QMNIST dataset facilitated evaluation of  instance level performance, and also allowed to observed a tendency of ABMIL to focus on, for the task, not relevant features.
We see a potential of PAP-QMNIST for usage in other similar evaluations.

\bibliography{sample}



\section*{Acknowledgements}
This work is supported by: Sweden’s Innovation Agency (VINNOVA), grants 2017-02447 and 2020-03611, and the Swedish Research Council, grant 2017-04385. A part of computations was enabled by resources provided by the Swedish National Infrastructure for Computing (SNIC) at Chalmers Centre for Computational Science and Engineering (C3SE), partially funded by the Swedish Research Council through grant no. 2018-05973.

\section*{Author contributions statement}
J.L., N.S. and N.K. conceived the experiments, N.K. conducted the experiments. V.B. shared expert knowledge of cytotechnologist and performed per-cell evaluation on cytological data. All authors reviewed the manuscript. 








\end{document}